\documentclass[11pt]{article} % arXiv preprint version: single column
\usepackage{ielabarxiv}       % house style: typography, panel, finding boxes

\usepackage[numbers,sort&compress]{natbib}
\usepackage{nicematrix}
\usepackage[normalem]{ulem}

\newcommand{\iter}{ITER}

\newcommand{\revisiondel}[1]{}
\newcommand{\revisionadd}[1]{#1}
\newcommand{\sigmark}[1]{\rlap{\hspace{1pt}\textsuperscript{#1}}}

\preprintheader{Preprint. Under review.}

\begin{document}

% ---- title panel content (rendered by ielabarxiv.sty) ----
\panelTitle{ITER: \\ Interaction-Aware Retrieval for Agentic Search}

\panelAuthors{%
 Haodong Chen\eqcontrib,\quad
 Shuai Wang\eqcontrib,\quad
 Yu Yin\par
 \vspace{0.25em}
 Shengyao Zhuang,\quad
 Guido Zuccon,\quad
 Teerapong Leelanupab}

\panelAffiliation{The University of Queensland, Brisbane, Australia}

\panelEmails{%
 haodong.chen1@student.uq.edu.au\par
 \{shuai.wang2, y.yin1, s.zhuang, g.zuccon, t.leelanupab\}@uq.edu.au}

\panelLinks{%
 \githublink{https://github.com/ielab/ITER}\par
 \hflink{https://huggingface.co/collections/ielabgroup/iter}}

\begin{titlepanel}
Deep-research agents answer complex user questions through an iterative sequence of search steps, where the agent autonomously formulates sub-queries to retrieve the evidence needed at each stage. However, existing retriever training typically relies only on the sub-query and its corresponding search results at the current step as training signals, leaving the information accumulated from previous interactions largely underutilized.

We introduce \iter{}, an agent interaction-aware dense retriever trained using agent trajectory learning signals. \iter{} represents each query by incorporating not only the current sub-query, but also the main question, \revisionadd{the agent's pre-search reasoning,} and preceding sub-queries, and is trained using trajectory-relative learning signals derived from the agent's interactions.

Across six agent backbones from three model families, \iter{} consistently outperforms the existing agent-trajectory-trained dense retriever, LRAT, achieving an average \revisionadd{relative }improvement of \revisiondel{7.5}\,\revisionadd{6.9}\% on InfoSeek-Eval and \revisiondel{13.5}\,\revisionadd{15.4}\% on BrowseComp-Plus. \revisiondel{\iter{} also demonstrates stronger cross-agent robustness than AgentIR, a deep-research retriever that relies on external LLM-judge signals and the agent's pre-search reasoning.} \revisionadd{At the matched 4B scale, \iter{} outperforms AgentIR on InfoSeek-Eval for five of six backbones while achieving a higher visit-to-search recall ratio on BrowseComp-Plus across all six backbones.} Ablations further show that \revisiondel{the main question and previous sub-queries provide the most robust query representation}\,\revisionadd{structured interaction history and pre-search reasoning provide complementary retrieval context}, while previously visited and useful documents, used as redundancy negatives in subsequent searches, provide the strongest trajectory-relative supervision.
\end{titlepanel}

\vspace{-0.3em}

% ---- teaser: overview figure on the title page ----
\begin{center}
  \begin{minipage}{\textwidth}
  \centering
  \includegraphics[width=1.0\textwidth]{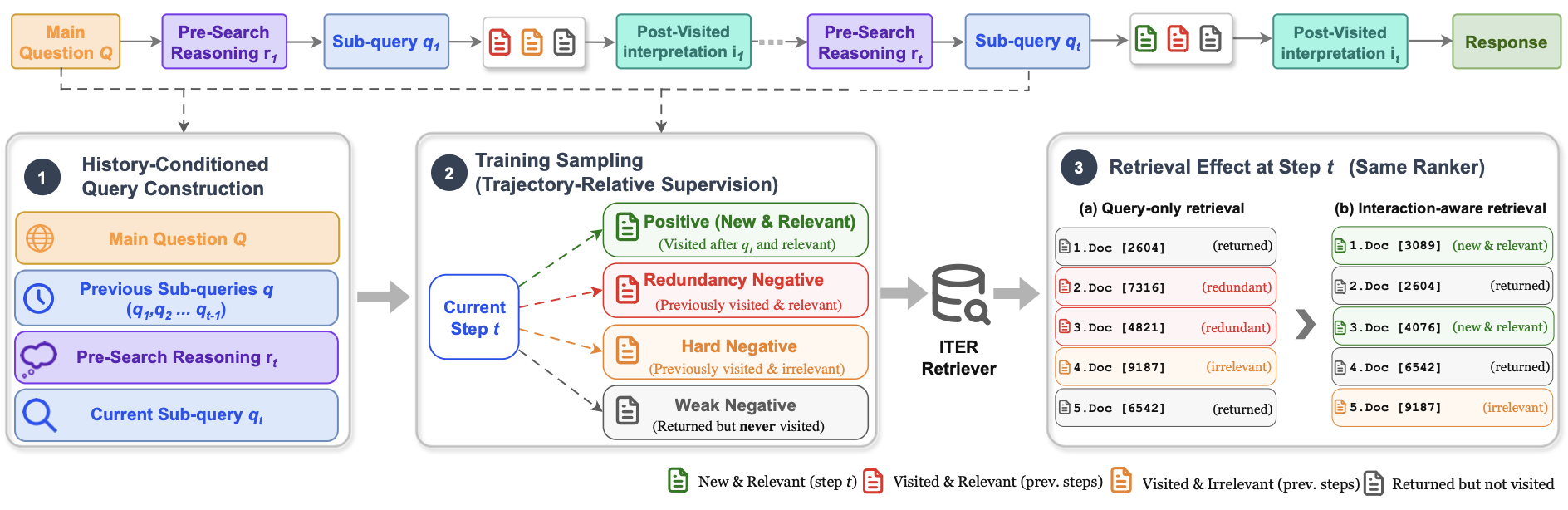}
  \captionsetup{font=footnotesize,skip=3pt}\captionof{figure}{Overview of \iter{}. \iter{} constructs a history-conditioned query
  from the main question, current pre-search reasoning, current sub-query, and
  previous sub-queries, and derives
  trajectory-relative supervision from the agent's document interactions. The
  resulting retriever promotes new relevant evidence while demoting previously
  visited redundant documents.}
  \label{fig:overview}
  \end{minipage}
\end{center}

\vspace{0.2em}

% \begin{figure}[t]
% \centering
% \includegraphics[width=\textwidth]{figures/example.png}

% \caption{Query-only vs.\ interaction-aware retrieval in an illustrative search
% trajectory. Ranking by the current query alone (a) re-surfaces candidates the
% agent has already visited (\textcolor{red}{red}) and misses the gold document.
% Conditioning the same search on the interaction history (b) promotes the unread
% gold document (\textcolor{green!55!black}{green}) and other supporting evidence
% (\textcolor{blue}{blue}), while retaining previously visited results at lower
% ranks. Results shared by both rankings are shown in gray.}
% \label{fig:example}
% \end{figure}

\begin{figure}[t]
\centering
\begin{tcolorbox}[colback=gray!6,colframe=black!55,boxrule=.35pt,
  left=3pt,right=3pt,top=2pt,bottom=2pt]
\scriptsize
\textbf{Main Question:} Which player from Kobe Bryant's 1996 draft class returned
to the team that originally drafted him and then won the NBA MVP award in
consecutive seasons?\\[1pt]
\textbf{Pre-Search Reasoning 1:} Kobe Bryant is the reference point in the
question, so I should first check whether he satisfies the MVP and return
conditions.\\
\textbf{Search 1:} \textit{["Kobe Bryant" "1996 NBA draft" "MVP" "guard"]}\\
\hspace*{1em}\textbf{Visit:} \texttt{Doc[4821]}: \emph{Kobe Bryant entered the
league in 1996 and won the 2008 MVP ...}\\
\hspace*{1em}\textbf{Document Interpretation:} Bryant belongs to the 1996 draft
class but won only one regular-season MVP.\\
\textbf{Pre-Search Reasoning 2:} Since Bryant won only one MVP, I should search
the wider 1996 draft class for an MVP who later returned to his original team.\\
\textbf{Search 2:} \textit{["1996 NBA draft" "MVP" "returned to original team"]}\\
\hspace*{1em}\textbf{Visit:} \texttt{Doc[7316]}: \emph{Allen Iverson won the 2001
MVP and returned to Philadelphia in 2009 ...}\\
\hspace*{1em}\textbf{Document Interpretation:} Iverson returned to the team that
drafted him but did not win MVP in consecutive seasons.\\
\textbf{Pre-Search Reasoning 3 (current):} Iverson also won only one MVP, so I
should narrow the remaining candidates to point guards who won consecutive MVPs.\\
\textbf{Search 3 (current):} \textit{["1996 NBA draft" "point guard" "MVP"
"returned to original team"]}
\end{tcolorbox}

\begin{minipage}[t]{0.487\textwidth}
\begin{tcolorbox}[colback=white,colframe=black!55,boxrule=.35pt,
  left=3pt,right=3pt,top=3pt,bottom=3pt,
  equal height group=teaserpanels,valign=top]
\scriptsize
\centerline{\textbf{(a) Query-only retrieval}}
\smallskip
\textbf{Retriever input:} \texttt{["1996 NBA draft" "point guard" "MVP"
"returned to original team"]}\\[2pt]
\textbf{Top-5 results:}\\[1pt]
\colorbox{red!8}{\parbox{\dimexpr\linewidth-2\fboxsep\relax}{\scriptsize 1.\ \texttt{Doc[7316]}: \emph{Iverson rejoined the 76ers in 2009 ...}\hfill
\textcolor{red!70!black}{[visited]}}}\\[1pt]
\colorbox{red!8}{\parbox{\dimexpr\linewidth-2\fboxsep\relax}{\scriptsize 2.\ \texttt{Doc[4821]}: \emph{Bryant won the 2008 MVP ...}\hfill
\textcolor{red!70!black}{[visited]}}}\\[1pt]
\colorbox{gray!10}{\parbox{\dimexpr\linewidth-2\fboxsep\relax}{\scriptsize 3.\ \texttt{Doc[2604]}: \emph{NBA MVP winners by year ...}\hfill
\textcolor{black!60}{[off target]}}}\\[1pt]
\colorbox{gray!10}{\parbox{\dimexpr\linewidth-2\fboxsep\relax}{\scriptsize 4.\ \texttt{Doc[9187]}: \emph{Notable stars of the 1996 draft ...}\hfill
\textcolor{black!60}{[off target]}}}\\[1pt]
\colorbox{gray!10}{\parbox{\dimexpr\linewidth-2\fboxsep\relax}{\scriptsize 5.\ \texttt{Doc[6542]}: \emph{Ten memorable NBA homecomings ...}\hfill
\textcolor{black!60}{[off target]}}}\\[2pt]
\emph{The gold document is not in the list.}\\[3pt]
\textcolor{red!70!black}{\ding{55}}~The agent finds no new evidence, falls back on
the candidate it already examined, and answers \emph{Allen Iverson} --- \textbf{wrong}.
\end{tcolorbox}
\end{minipage}\hfill
\begin{minipage}[t]{0.487\textwidth}
\begin{tcolorbox}[colback=white,colframe=black!55,boxrule=.35pt,
  left=3pt,right=3pt,top=3pt,bottom=3pt,
  equal height group=teaserpanels,valign=top]
\scriptsize
\centerline{\textbf{(b) Interaction-aware retrieval (\iter{}, ours)}}
\smallskip
\textbf{Retriever input:} \texttt{Main Question} $+$
\texttt{Current Pre-Search Reasoning} $+$ \texttt{Current Sub-query} $+$
\texttt{Previous Sub-queries}\\[2pt]
\textbf{Top-5 results:}\\[1pt]
\colorbox{green!12}{\parbox{\dimexpr\linewidth-2\fboxsep\relax}{\scriptsize 1.\ \texttt{Doc[3089]}: \emph{Nash won consecutive MVPs ...}\hfill
\textcolor{green!40!black}{[new, gold]}}}\\[1pt]
\colorbox{blue!8}{\parbox{\dimexpr\linewidth-2\fboxsep\relax}{\scriptsize 2.\ \texttt{Doc[5774]}: \emph{Nash returned to the Suns in 2004 ...}\hfill
\textcolor{blue!60!black}{[new]}}}\\[1pt]
\colorbox{gray!10}{\parbox{\dimexpr\linewidth-2\fboxsep\relax}{\scriptsize 3.\ \texttt{Doc[2604]}: \emph{NBA MVP winners by year ...}\hfill
\textcolor{black!60}{[off target]}}}\\[1pt]
\colorbox{gray!10}{\parbox{\dimexpr\linewidth-2\fboxsep\relax}{\scriptsize 4.\ \texttt{Doc[9187]}: \emph{Notable stars of the 1996 draft ...}\hfill
\textcolor{black!60}{[off target]}}}\\[1pt]
\colorbox{red!8}{\parbox{\dimexpr\linewidth-2\fboxsep\relax}{\scriptsize 5.\ \texttt{Doc[7316]}: \emph{Iverson rejoined the 76ers in 2009 ...}\hfill
\textcolor{red!70!black}{[visited]}}}\\[2pt]
\emph{The unread gold document is promoted, while previously visited candidates rank lower.}\\[3pt]
\textcolor{green!40!black}{\ding{51}}~The agent opens the Steve Nash document and
answers \emph{Steve Nash} --- \textbf{correct}.
\end{tcolorbox}
\end{minipage}
\caption{Query-only vs.\ interaction-aware retrieval in an illustrative search
trajectory. Ranking by the current query alone (a) re-surfaces candidates the
agent has already visited (\textcolor{red}{red}) and misses the gold document.
Conditioning the same search on the current pre-search reasoning and interaction
history (b) promotes the unread gold document
(\textcolor{green!55!black}{green}) and other supporting evidence
(\textcolor{blue}{blue}), while retaining previously visited results at lower
ranks. Results shared by both rankings are shown in gray.}
\label{fig:example}
\end{figure}

\section{Introduction}
\label{sec:intro}
Search has long benefited from interaction signals. In web search, relevance feedback and click logs reveal which results users find helpful, providing supervision for improving retrieval through techniques such as query reformulation, query expansion, and learning to rank~\cite{rocchio1971,lavrenko2001,agichtein2006,joachims2002,joachims2005}. Agentic search produces a new, more structured form of interaction within the search loop. An agent repeatedly issues sub-queries, examines retrieved documents, and revises what to search for next as evidence accumulates and its understanding of the task evolves~\cite{white2024,tongyi2025,webthinker2025}. Each agent execution leaves a trajectory recording what the agent searched for, which documents it examined, and what it learned from them.

Recent work has begun training retrievers for agentic search using signals derived from agent interactions, such as an agent’s decision to visit a full document or its reasoning before issuing a sub-query~\cite{lrat2026,agentir2026}. However, these approaches largely treat each search step in isolation, without fully accounting for previous sub-queries or previously visited documents.
This omission matters for two reasons. First, previous sub-queries reveal which search directions the agent has already explored, helping the retriever recognize sub-query reformulations and avoid resurfacing the same results. Second, previously visited documents indicate which parts of the information need may already have been satisfied, allowing the retriever to prioritize evidence that addresses what remains unresolved. 
Therefore, a document’s utility should depend not only on its relevance to the current sub-query, but also on the new information it contributes beyond the evidence already collected. We refer to this history-dependent utility as the \emph{marginal gain} of the document. 
Figure~\ref{fig:example} illustrates how a current-sub-query-only retriever can resurface previously visited candidates while overlooking an unvisited document that addresses the remaining information need.

To model this interaction-dependent document utility, we formulate \emph{interaction-aware retrieval}, in which document ranking is conditioned on the agent’s interaction history rather than on the current sub-query alone. We introduce the \textbf{I}nteraction-aware \textbf{T}rajectory-conditioned \textbf{E}mbedding \textbf{R}etriever (\iter{}), an agent-trajectory-trained dense retriever for this setting. 
At each search step, \iter{} constructs a history-conditioned query representation by combining the main question, \revisionadd{current pre-search reasoning,} current sub-query, and previous sub-queries. The main question preserves the overall task, \revisionadd{the current pre-search reasoning and} current sub-query specify the immediate information need, and the previous sub-queries indicate the search directions already explored. \iter{} learns from agent trajectories collected in a de-duplicated search setting, using document interactions to construct step-specific positives and tiered negatives: documents visited after the current search and judged relevant are positives; previously visited relevant and irrelevant documents are redundancy and hard negatives, respectively; and current results never visited in the trajectory are weak negatives. Figure~\ref{fig:overview} provides an overview of \iter{}'s history-conditioned query representation, trajectory-relative training supervision, and resulting retrieval behaviour.

%such as which documents the agent opens, or from the agent’s reasoning at the current search step~\cite{lrat2026,agentir}. However, this supervision is still derived separately at each step. Even when previous subqueries are included as context, positive and negative documents are selected from the current step alone. The resulting signals therefore do not capture whether a document provides new information or merely repeats evidence the agent has already obtained.

%Search agents naturally produce these trajectories, which can also be used to improve the retriever. Recently, LRAT~\cite{lrat2026} explored how agent search and browsing actions can provide effective training signals for retrieval. In LRAT, however, each document is still evaluated with respect to the current query, without considering whether the agent has already consumed its information earlier in the trajectory.

We evaluate \iter{} on InfoSeek-Eval~\cite{infoseek2025} and BrowseComp-Plus~\cite{browsecompplus2025}. When evaluated with Tongyi-DeepResearch-30B, whose trajectories are used for training, the default \revisionadd{0.6B }\iter{} achieves a task success rate of \revisiondel{80.0}\,\revisionadd{78.7} on InfoSeek-Eval, compared with \revisiondel{72.7}\,\revisionadd{72.0} for LRAT and \revisiondel{76.7}\,\revisionadd{72.0} for the variant using only the current sub-query. On BrowseComp-Plus, it achieves \revisiondel{46.6}\,\revisionadd{49.2}, compared with \revisiondel{43.4}\,\revisionadd{43.7} and \revisiondel{43.7}\,\revisionadd{45.3}, respectively. Across six agent backbones from three model families, \revisionadd{the 0.6B }\iter{} outperforms LRAT in all 12 backbone--benchmark comparisons\revisiondel{, with seven statistically significant gains}. \revisiondel{On the five unseen agent backbones, \iter{} also achieves higher task success than AgentIR on both benchmarks.} \revisionadd{At the matched 4B scale, \iter{} achieves higher task success than AgentIR on InfoSeek-Eval for five of six agent backbones and a higher visit-to-search recall ratio on BrowseComp-Plus across all six backbones.} Ablations show that the main question and previous sub-queries provide \revisiondel{the most}\,\revisionadd{a} robust query representation, while pre-search reasoning \revisiondel{transfers less consistently}\,\revisionadd{provides complementary retrieval context} and adding visited documents or their interpretations reduces evidence search recall. Redundancy negatives provide the strongest trajectory-relative supervision signal.

Our contributions are as follows:
%\vspace{-4pt}
\begin{itemize}[leftmargin=*]
\item We formulate interaction-aware retrieval for deep-research agents, where a document's utility reflects its marginal gain given the current sub-query and interaction history, rather than its relevance to the current sub-query alone.

\item We introduce \iter{}, which combines a history-conditioned query with trajectory-relative supervision comprising step-specific positives and tiered negatives derived from agent trajectories.

\item Through matched-agent and cross-agent evaluations, we show that \revisiondel{\iter{} consistently outperforms LRAT and exhibits stronger cross-agent robustness than AgentIR}\,\revisionadd{interaction-aware training improves task performance across diverse agents and better aligns retrieval with the evidence that agents actually consume}; ablations further reveal how interaction signals and negative tiers contribute to these gains.

\end{itemize}
   % Introduction
\section{Related Work}
\label{sec:related}

\subsection{Agentic and Deep-Research Search}

Retrieval-augmented generation supplements language-model generation with evidence
retrieved from external corpora~\cite{rag2020}, but early RAG pipelines typically
retrieve once for the input question. Tool-using systems such as WebGPT, ReAct, and
IRCoT turn search into an action within the problem-solving process: the model can
retrieve evidence, examine what it finds, and search again
~\cite{webgpt2021,react2023,ircot2023}. Recent deep-research agents build on this
iterative pattern to address questions that require extended exploration across
multiple sources~\cite{searcho12025,searchr12025,deepresearcher2025,drsurvey2025}. They
formulate a sequence of sub-queries, accumulate evidence across retrieval steps, and
eventually synthesize the collected information into an answer or report. We study
the retriever that supports these successive search requests.

\subsection{Optimizing Retrieval for Agentic Search}

Search agents often rely on retrievers optimized for stand-alone query--document
matching~\cite{dpr2020,ance2021}. In a multi-step trajectory, however, each query arises from the agent's
evolving search process and serves an intermediate information need. Recent work
adapts retrieval to this setting. NExT-Search emphasizes feedback at intermediate
stages of generative search~\cite{nextsearch2025}, while Agentic-R trains retrievers
using local query-passage relevance and global answer correctness~\cite{agenticr2026}. LRAT
converts the agent's search and browsing actions into supervision for the retriever
~\cite{lrat2026}. Other studies examine how retrieval granularity, ranking models,
query transformations, and lexical or dense matching affect deep-research agents
    ~\cite{meng2026ranking,sage2026}. \iter{} belongs to this line of work, but defines
document utility relative to the interaction state in which each search occurs.

A second direction changes how the agent accesses retrieved evidence. Direct Corpus
Interaction lets the agent operate on the collection with shell commands instead of
receiving only a ranked candidate list~\cite{dci2026}. RISE uses retrieval to form a
bounded workspace that the agent can continue to inspect~\cite{rise2026}. SIEVE changes corpus access by combining fielded Boolean queries with structure-aware result inspection and selective section fetching~\cite{sieve2026}. These systems
optimize the retrieval workflow or corpus interface. \iter{} instead retains a
ranked-retrieval interface and optimizes the retriever behind it.

\subsection{Context-Conditioned Retrieval}

Context-conditioned retrieval uses information beyond the current query to guide
retrieval. The additional context may come from the task itself: task-aware
retrievers condition on instructions that specify the desired evidence
~\cite{tart2023}. In conversational search, context accumulates through interaction.
Systems either rewrite the latest utterance using earlier turns~\cite{cqr2020} or
encode the utterance and dialogue history together
~\cite{convdr2021,haconvdr2024,chatretriever2024,contextualretriever2025}. Context can also be generated: HyDE
constructs hypothetical content and uses it as an intermediate retrieval
representation~\cite{hyde2023}. In each case, additional context helps the retriever
better identify what information is needed at the current step.

Deep-research trajectories provide interaction context from the agent's own search
process. AgentIR augments the current sub-query with pre-search reasoning to provide
additional context for retrieval~\cite{agentir2026}. In \iter{}, \revisiondel{interaction context
serves a different purpose: rather than expanding the current sub-query, it}\,\revisionadd{pre-search reasoning is combined with the main question and previous sub-queries to form structured interaction context that} anchors
retrieval to the agent's current search state, reflecting what has already been
explored.
   % Related Work
\section{Observations from Agent Search}
\label{sec:method:observation}

\begin{figure}[t]
\centering
\includegraphics[width=\textwidth]{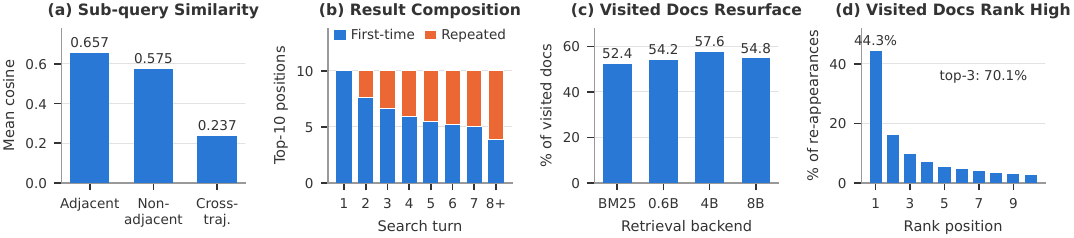}
\caption{Agent search behaviour on the 26{,}482 trajectories released by
LRAT (Tongyi-DeepResearch, standard top-10 retrieval setting,
four retrieval backends). (a)~Adjacent sub-queries remain semantically close
under a dense encoder that receives only the current sub-query. (b)~Repeated
documents occupy an increasing share of the ranking as the trajectory
progresses. (c)~Most visited documents are returned again by later searches.
(d)~When they reappear, they concentrate at the top of the ranking.}
\label{fig:traj-stats}
\end{figure}

To identify useful interaction signals for retrieval, we analyze the 26{,}482 agent trajectories released by LRAT~\cite{lrat2026}, collected with Tongyi-DeepResearch across four retrieval backends in a standard top-10 retrieval setting. Figure~\ref{fig:traj-stats} reports sub-query similarity, result novelty, and the recurrence of visited documents over successive searches.

\finding{1}{Query reformulation alone does not prevent repeated results.}

Agents typically refine an earlier sub-query to address a remaining information gap. As shown in Figure~\ref{fig:traj-stats}(a), adjacent sub-queries have a mean cosine similarity of .657. Non-adjacent sub-queries from the same trajectory also remain similar (.575), whereas random cross-trajectory pairs average only .237. This continuity extends to the rankings: the number of first-time documents in the top 10 steadily decreases from 10.0 at the first search to 7.6 at the second and 6.6 at the third, and continues to decline as the trajectory progresses. From the eighth search onward, only 3.9 out of 10 results are new on average (Figure~\ref{fig:traj-stats}(b)). Previous sub-queries therefore reveal which search directions have already been explored and can help retrieval prioritize new evidence.

\finding{2}{Previously visited documents frequently resurface at the top.}

However, recurrence alone does not establish redundancy. A returned but unvisited document may have been overlooked or deferred and may still be useful. A visited document has already been examined and, if judged relevant, has contributed useful evidence. Across the four retrieval backends, on average  54.9\% of visited documents appear again in a subsequent search (Figure~\ref{fig:traj-stats}(c)). Of these reappearances, 44.3\% occur at rank 1 and 70.1\% within the top three positions (Figure~\ref{fig:traj-stats}(d)). A retriever that sees only the current sub-query cannot make this distinction, making previously visited relevant documents a natural redundancy signal for subsequent searches.

\subsection{Interaction-Aware Retrieval}

Together, these observations reveal a mismatch between multi-step agent search and conventional retrievers that process each search query independently. As the agent examines documents, some information needs are resolved while others remain. Yet the retriever sees only the latest sub-query, not the agent's earlier queries or document visits, and may therefore rank previously examined documents above documents containing new evidence.

\emph{Interaction-aware retrieval} addresses this mismatch by ranking documents using \revisiondel{both }the current sub-query, \revisionadd{current pre-search reasoning,} and the interaction history. Document utility is therefore \emph{interaction-dependent}. At step $t$, the utility of document $d$ is its marginal gain,
\[
g_t(d) = \operatorname{gain}(d \mid q_t, \revisionadd{\tau_t,} H_t),
\]
rather than its relevance to $q_t$ alone. Once a document has been visited, its utility may decrease even if it remains topically relevant.

To define the interaction history, consider a trajectory at search step $t$. The agent \revisionadd{produces pre-search reasoning $\tau_t$ and then} issues a sub-query $q_t$, retrieves $k$ documents with short snippets, and may visit selected documents before searching again. Before issuing $q_t$, it has the history
\[
H_t =
\left(
Q,
\{q_i\}_{i<t},
R_{<t},
V_{<t},
I_{<t}
\right),
\]
where $Q$ is the main question; $\{q_i\}_{i<t}$ contains the previous sub-queries; $R_{<t}$ and $V_{<t}$ denote the documents retrieved and visited before step $t$, respectively; and $I_{<t}$ contains document-specific interpretations extracted from post-visit reasoning, summarizing what the agent learned from each visited document.

This formulation motivates two design choices for \iter{}. First, the retriever receives \revisionadd{current pre-search reasoning and} context from earlier searches. Second, document interactions provide training signals that reflect a document's utility at each search step.
% Interaction-aware retrieval instead aims to select more useful evidence within
% overlapping retrieval regions, rather than simply increasing retrieval novelty.

% Because $g_t(d)$ is latent, \iter{} uses agent interactions as state-relative
% supervision. A document first visited after $q_t$ and judged useful becomes a positive
% under $H_t$, whereas a useful document visited before step $t$ becomes a redundancy
% negative once its information has been consumed. Section~\ref{sec:method:examples}
% describes how these signals are constructed.
   % Revisiting Agent Search Behaviour
\section{\iter{}: Query Representation and Training}
\label{sec:method}

\iter{} implements the two design choices motivated in Section~\ref{sec:method:observation}: conditioning retrieval on \revisionadd{current pre-search reasoning and} the interaction history and deriving trajectory-relative supervision from document interactions.

\subsection{History-Conditioned Query Representation}
\label{sec:method:query}

The interaction history contains several sources of information that may guide retrieval, including the main question, previous sub-queries, visited documents, and their associated interpretations. \iter{} incorporates this history by serializing selected components together with the current sub-query as input to the query encoder.

The default query representation combines the main question $Q$, \revisionadd{current pre-search reasoning $\tau_t$,} current sub-query $q_t$, and previous sub-queries $q_{<t}=\{q_i\}_{i<t}$. The main question preserves the overall task, \revisionadd{the current pre-search reasoning and} the current sub-query \revisiondel{specifies}\,\revisionadd{specify} the immediate information need, and the previous sub-queries indicate the search directions already explored. At the first search step, $q_{<t}$ is empty. \revisiondel{AgentIR augments the current sub-query with pre-search reasoning~\cite{agentir2026}. To examine whether pre-search reasoning provides additional context, we also consider a reasoning-augmented representation that adds $\tau_t$, the reasoning produced before issuing $q_t$, together with the corresponding instruction text.}\ \revisionadd{Here, $\tau_t$ is the reasoning produced before issuing $q_t$ and, following AgentIR~\cite{agentir2026}, is serialized together with reasoning-specific instruction text. Figure~\ref{fig:history-query} illustrates the resulting query representation.} The effects of visited documents, document interpretations, and other combinations of interaction history are examined in the query-representation ablation.

% ========================================== figure 3 ==========================================

\begin{figure}[t]
\centering
\begin{tcolorbox}[colback=gray!4,colframe=black!55,boxrule=.35pt,
  left=3pt,right=3pt,top=3pt,bottom=1pt]
\small
\textbf{Instruction:} Given the main question,
\textcolor{blue!70!black}{the agent's reasoning and} the current sub-query
\textcolor{blue!70!black}{it led to}, and the
sub-queries already tried in previous interactions, retrieve documents relevant to
the current sub-query that provide new information not yet found.\\[2pt]
\textbf{Main Question:} Which player from Kobe Bryant's 1996 draft class returned
to the team that originally drafted him and then won the NBA MVP award in
consecutive seasons?\\
\textcolor{blue!70!black}{\textbf{\revisiondel{Current Reasoning}\,\revisionadd{Current Pre-Search Reasoning}:} Kobe Bryant and Allen
Iverson were both selected in 1996, but each won only one regular-season MVP. We
need another point guard from that draft class who returned to his original team
and won the award in consecutive seasons.}\\
\textbf{Current Sub-query:} ["1996 NBA draft" "point guard" "MVP" "returned to
original team"]\\
\textbf{Previous Interactions:}\\
\hspace*{.5em}\textbf{Previous Sub-query 1:} ["Kobe Bryant" "1996 NBA draft"
"MVP" "guard"]\\
\hspace*{.5em}\textbf{Previous Sub-query 2:} ["1996 NBA draft" "MVP" "returned
to original team"]
\end{tcolorbox}
\caption{\revisiondel{Default and reasoning-augmented query representations. Both use the
content shown in black; the reasoning-augmented representation additionally includes the
pre-search reasoning and instruction text shown in blue.}\ \revisionadd{The default \iter{}
query representation uses all content shown. Omitting pre-search reasoning removes the
reasoning and reasoning-specific instruction text shown in blue.} The example is truncated for display.}
\label{fig:history-query}
\end{figure}
% ========================================== figure 3 ==========================================

\begin{figure}[t]
\centering
\begin{tcolorbox}[colback=gray!4,colframe=black!55,boxrule=.35pt,
  left=3pt,right=3pt,top=3pt,bottom=3pt]
\small
\textcolor{blue!65!black}{$d$}~first-time result \quad
\textcolor{red!70!black}{$d$}~returned earlier \quad
\textcolor{green!45!black}{$d$}~promoted from below top-$k$\\[3pt]
\textbf{(a) Standard search results}\\
\hspace*{.5em}Step 1:
[\,\textcolor{blue!65!black}{$d_1\;d_2\;d_3\;d_4\;d_5$}\,\ldots]\\
\hspace*{.5em}Step 2:
[\,\textcolor{red!70!black}{$d_2$}\;\textcolor{blue!65!black}{$d_6$}\;
\textcolor{red!70!black}{$d_1$}\;\textcolor{blue!65!black}{$d_7$}\;
\textcolor{red!70!black}{$d_3$}\,\ldots]\\
\hspace*{.5em}\emph{Three of five positions repeat earlier results.}\\[4pt]
\textbf{(b) De-duplicated search results}\\
\hspace*{.5em}Step 2 top-$k$:
[\,\textcolor{blue!65!black}{$d_6\;d_7$}\;
\textcolor{green!45!black}{$d_8\;d_9\;d_{10}$}\,\ldots]\\
\hspace*{.5em}\texttt{returned earlier:}
\textcolor{red!70!black}{$d_1,\,d_2,\,d_3$}\\
\hspace*{.5em}\emph{Every ranked position is new; earlier results remain openable
via \texttt{get\_document}.}
\end{tcolorbox}
\caption{Standard and de-duplicated search settings. Under standard retrieval,
previously returned documents occupy positions in the next result list. The
de-duplicated interface fills those positions with unseen documents and moves the
repeated documents to a \texttt{returned earlier} section, where they remain
accessible through the \texttt{get\_document} tool.}
\label{fig:dedup-example}
\end{figure}

\subsection{Constructing Training Signals from Agent Trajectories}
\label{sec:method:signals}

Agent trajectories record when documents are returned, when they are visited, and whether they contribute useful information. The timing of these interactions is important: a document visited and judged relevant after search $t$ provides a positive signal for that search, but may indicate redundancy in later searches. Conversely, a document not visited when first returned may still prove useful later. We therefore collect trajectories in a de-duplicated search setting and use the resulting document interactions to construct positive pairs and tiered negatives.

\subsubsection{De-duplicated Trajectory Collection}
\label{sec:method:harness}

Successive sub-queries often return the same highly ranked documents, leaving fewer positions for unseen candidates. To increase candidate coverage without removing access to earlier results, we use a de-duplicated search setting that separates first-time results from previously returned documents.

At each search step, the search tool retrieves $K{=}100$ candidates, removes documents returned at earlier steps, and presents the top $k{=}10$ unseen documents. If a previously returned document appears in the raw top-10 results for the current sub-query, its identifier and title are displayed separately under \texttt{returned earlier}. The agent can still open the document using \texttt{get\_document}. Figure~\ref{fig:dedup-example} contrasts the standard and de-duplicated settings.

We run Tongyi-DeepResearch-30B~\cite{tongyi2025} on 10{,}000 InfoSeek training questions~\cite{infoseek2025} with four retrieval backends: BM25~\cite{bm25} and the zero-shot Qwen3-Embedding models~\cite{qwen3embedding2025} at 0.6B, 4B, and 8B scales. This produces 40{,}000 trajectories with different candidate distributions. We retain the 20{,}893 trajectories whose final answers match the references.

The de-duplicated setting presents more unseen candidates while preserving access to earlier results. Among the 67{,}934 positive pairs derived from the retained trajectories, 18{,}138 (26.7\%) involve delayed visits, where the agent visits a document returned by an earlier search. These delayed visits show that not visiting a document when it first appears does not establish that it is unhelpful.

\subsubsection{Positive Signals from Document Visits}

For each search step $t$, we pair the history-conditioned query representation for $q_t$ with documents visited after $q_t$ and before the next search. A visited document may come from either the current results or \texttt{returned earlier}. If a document was returned at an earlier step but visited only after $q_t$, it is paired with $q_t$ and the corresponding interaction history, rather than with the sub-query that first returned it.

A visit alone does not establish relevance because the document may prove unhelpful after the agent examines its full content. Following LRAT~\cite{lrat2026}, we provide the reasoning produced after each visit to Qwen3-30B-A3B-Thinking-2507~\cite{qwen3}, which judges whether the document contributed useful information. The verifier outputs \textsc{Relevant} or \textsc{Not Relevant}, and only documents judged \textsc{Relevant} are retained as positives.

\subsubsection{Negative Signals from Document Interactions}
\label{sec:method:negatives}

For each positive pair at search step $t$, \iter{} organizes negative documents into three tiers according to the agent's interactions with them:

\begin{itemize}[leftmargin=*]
\item \textbf{Redundancy negatives:} documents visited before step $t$ and judged relevant. They remain topically related but have already provided useful information to the agent.

\item \textbf{Hard negatives:} documents visited before step $t$ and judged irrelevant. Their snippets appeared useful enough to prompt a visit, but their full content was judged unhelpful.

\item \textbf{Weak negatives:} documents returned at step $t$ but never visited anywhere in the complete trajectory. Because the agent never examined them, their utility remains uncertain.
\end{itemize}

Documents returned only at earlier steps and never visited remain unlabeled. As the delayed visits demonstrate, the absence of an immediate visit does not establish that a document is unhelpful.

Each training group contains one positive and nine negatives. We sample up to three redundancy negatives, followed by up to three hard negatives, and fill the rest with weak negatives.

\subsection{Trajectory-Relative Training}
\label{sec:method:training}

Given the constructed training groups, \iter{} is optimized with a weighted contrastive objective. The objective uses two forms of weighting: an instance-level weight derived from the reasoning associated with each positive document and a tier-specific weight for each sampled negative.

\subsubsection{Positive-Instance Weighting}

Relevance filtering provides a binary label, but the reasoning associated with a positive document can provide a graded signal of its utility. LRAT observes that longer post-visit reasoning traces are associated with more useful documents~\cite{lrat2026}. Following LRAT, we apply the same length-to-weight mapping.

Let $\ell_i$ denote the token length of the post-visit reasoning associated with the positive document in training instance $i$, and let $\beta$ be the median positive reasoning length in the training set. We compute the saturating raw score
\begin{equation}
\label{eq:positive-raw-weight}
\tilde{a}_i =
1-\exp\!\left(-\frac{\ln 2}{\beta}\ell_i\right),
\end{equation}
and normalize it to have mean one:
\begin{equation}
\label{eq:positive-weight}
a_i =
\frac{\tilde{a}_i}{\mathbb{E}_j[\tilde{a}_j]}.
\end{equation}

The raw score reaches half of its asymptotic value at $\ell_i=\beta$ and gradually saturates for longer reasoning traces. The resulting $a_i$ controls the contribution of training instance $i$ to the final objective.

\subsubsection{Negative-Tier Weighting}

The negative tiers defined in Section~\ref{sec:method:negatives} provide different levels of evidence that a document should be ranked below the current positive. We therefore assign them different weights: $(w_{\mathrm{red}},w_{\mathrm{hard}},w_{\mathrm{weak}})
=
(3.0,1.0,0.3)$.
% \[
% (w_{\mathrm{red}},w_{\mathrm{hard}},w_{\mathrm{weak}})
% =
% (3.0,1.0,0.3).
% \]

Redundancy negatives receive the largest weight because they were previously judged useful but have already provided information to the agent. Hard negatives receive the standard weight, while weak negatives are discounted because their utility was never directly observed. Table~\ref{tab:abl-neg} evaluates the effects of the negative tiers and their weights.

\subsubsection{Weighted Contrastive Loss}

For a mini-batch of history-conditioned query representations $q_i$, positive documents $d_i^+$, and sampled negative groups $\mathcal{G}_i$, documents from other training instances also serve as in-batch negatives. Let $\mathcal{B}$ contain all documents in the batch, let $s(\cdot,\cdot)$ denote cosine similarity, and let $\tau=0.02$ be the softmax temperature.

For each training instance $i$, we define
\begin{equation}
\label{eq:instance-loss}
\mathcal{L}_i =
-\log
\frac{e^{s(q_i,d_i^+)/\tau}}
{\displaystyle\sum_{d\in\mathcal{B}}
w_i(d)e^{s(q_i,d)/\tau}},
\end{equation}
where $w_i(d)$ takes the corresponding tier weight when $d\in\mathcal{G}_i$ and is one for the positive document and all other in-batch documents. Multiplying a negative term by $w_i(d)$ is equivalent to adding $\log w_i(d)$ to its logit, thereby controlling how strongly the negative competes with the positive. The instance weights $a_i$ from Eq.~\ref{eq:positive-weight} are then used to aggregate the per-instance losses:
\begin{equation}
\label{eq:loss}
\mathcal{L} =
\frac{\sum_i a_i\mathcal{L}_i}
{\sum_i a_i}.
\end{equation}

Thus, $w_i(d)$ controls how strongly each sampled negative competes with the positive within an instance, while $a_i$ controls the contribution of the entire instance to the final objective.

\subsubsection{Training Parameters}

We implement retriever training with the FlagEmbedding framework~\cite{flagembedding2023}. We fully fine-tune Qwen3-Embedding-0.6B~\cite{qwen3embedding2025}, the backbone used by LRAT, for two epochs using AdamW with a learning rate of $10^{-6}$, a 0.1 warmup ratio, batch size 32, bf16 precision, last-token pooling, and normalized embeddings. Documents are truncated to 512 tokens, while query inputs are truncated to 8{,}192 tokens to accommodate the alternative interaction inputs. We use the same training procedure for the 4B scaling experiment.
   % ITER
\section{Experimental Setup}
\label{sec:setup}

\subsection{Evaluation Benchmarks}

We evaluate \iter{} on two fixed-corpus deep-research benchmarks using the same in-domain and out-of-domain setup as LRAT~\cite{lrat2026}.

\textbf{InfoSeek-Eval}~\cite{infoseek2025} contains 300 multi-hop information-seeking questions strictly disjoint from the InfoSeekQA questions used for trajectory collection. Retrieval uses Wiki-25-Dump\footnote{\url{https://huggingface.co/datasets/Lk123/wiki-25-512}}, which contains 11.2 million documents of up to 512 tokens. Because training and evaluation share the task distribution and retrieval corpus, but not questions, we treat InfoSeek-Eval as the in-domain benchmark.

\textbf{BrowseComp-Plus}~\cite{browsecompplus2025} is a reproducible benchmark derived from BrowseComp~\cite{browsecomp2025} and designed for deep-research agents. It contains 830 complex, human-authored questions requiring multi-step reasoning and evidence aggregation, often through long search trajectories. Following the official setup, retrieval uses a corpus of 100{,}195 documents. The benchmark provides document-level \emph{gold} and \emph{evidence} qrels for retrieval evaluation. Neither its questions nor its corpus is used during \iter{} training, making it our out-of-domain benchmark.

\subsection{Evaluation Metrics}

We report three groups of metrics. End-to-end effectiveness is measured by \textbf{task success rate (SR)}, the fraction of questions answered correctly. We use Qwen3-30B-A3B-Thinking-2507~\cite{qwen3} as the LLM judge for BrowseComp-Plus and exact string matching for InfoSeek-Eval.

Retrieval quality on BrowseComp-Plus is measured using two metrics. \textbf{Evidence search recall} is the fraction of annotated evidence documents retrieved at any point in an agent trajectory. \textbf{Evidence visit recall} is the fraction of those documents that the agent visits through the \texttt{get\_document} tool. Both are macro-averaged over questions. Execution efficiency is measured by the average number of tool calls per question (Avg. Steps). For paired comparisons over the same question set, we report exact McNemar $p$-values on per-question answer outcomes~\cite{mcnemar1947}.

\subsection{Compared Retrievers}

We compare \iter{} against sparse and dense retrievers. The dense retrievers use the Qwen3-Embedding family~\cite{qwen3embedding2025}. We use the 0.6B encoder for comparisons with LRAT and the 4B encoder \revisionadd{with the default representation} to study retriever scaling and compare with AgentIR at the same scale:

\begin{itemize}[leftmargin=*]
\item \textbf{BM25}~\cite{bm25}: a sparse lexical baseline without training, queried with the current sub-query alone;
\item \textbf{Base}: the pretrained Qwen3-Embedding model without trajectory fine-tuning, queried with the current sub-query alone;
\item \textbf{LRAT}~\cite{lrat2026}: a trajectory-trained retriever queried with the current sub-query alone. We use its released Qwen3-Embedding-0.6B checkpoint; LRAT does not provide a 4B model;
\item \textbf{AgentIR}~\cite{agentir2026}: the officially released AgentIR-4B model, evaluated with its reasoning-aware query representation consisting of the pre-search reasoning, current sub-query, and instruction prefix;
\item \textbf{\iter{}} (ours): a retriever trained with the trajectory-relative supervision described in Section~\ref{sec:method:signals}. We evaluate three query representations: the default representation combining the main question, \revisionadd{pre-search reasoning,} current sub-query, and previous sub-queries; a diagnostic representation using only the current sub-query; and a \revisiondel{reasoning-augmented representation that adds pre-search reasoning to the default}\,\revisionadd{second diagnostic representation that omits pre-search reasoning from the default}. These representations are described in Section~\ref{sec:method:query}.
\end{itemize}

During evaluation, all retrievers return unfiltered, \revisionadd{non-de-duplicated} top-$k$ rankings through the same search tool.

\subsection{Agent Backbones}

We use Tongyi-DeepResearch-30B~\cite{tongyi2025}, which was used for trajectory collection, in the matched-agent evaluation. To assess cross-agent transfer, we additionally evaluate five unseen backbones from two other model families, ranging from 4B to 120B parameters: Qwen3.5-4B/9B/27B, Qwen3.6-27B~\cite{qwen35}, and gpt-oss-120B~\cite{gptoss2025}.

\subsection{Implementation Details}

All agent backbones are served locally with vLLM~\cite{vllm2023}. We use the recommended generation settings for each backbone and keep them fixed across retrievers. At each search step, every retriever returns the top-10 documents with 64-token snippets. Agents can access full documents through the \texttt{get\_document} tool and are limited to 50 tool-calling turns per question.
   % Experimental Setup

\begin{table}[htbp]
\centering
\caption{Matched-agent evaluation with Tongyi-DeepResearch-30B. SR = task success rate;
Search/Visit Recall = BrowseComp-Plus evidence search/visit recall; Steps = mean
tool calls. MQ/SQ/PSQ/PR = main question, current sub-query, previous sub-queries,
pre-search reasoning. SR and recall annotations in the 0.6B \iter{} block show
relative changes over LRAT (\textcolor{red}{red}: increase).
Stars indicate significant differences from LRAT under two-sided exact McNemar
tests: $^{*}p{<}.05$.}
\label{tab:main}
\small
\renewcommand{\arraystretch}{1.2}
\setlength{\tabcolsep}{5pt}
\resizebox{\textwidth}{!}{%
\begin{tabular}{l l c c c c c c}
\toprule
\multirow{2}{*}{Method} & \multirow{2}{*}{Configuration}
  & \multicolumn{2}{c}{\textbf{InfoSeek-Eval} (ID)}
  & \multicolumn{4}{c}{\textbf{BrowseComp-Plus} (OOD)} \\
\cmidrule(lr){3-4}\cmidrule(lr){5-8}
& & SR ($\uparrow$) & Steps ($\downarrow$)
  & SR ($\uparrow$) & Search Recall ($\uparrow$) & Visit Recall ($\uparrow$)
  & Steps ($\downarrow$) \\
\midrule
\multirow{3}{*}{Baselines}
& BM25 (sparse) & 77.3 & 17.4 & 31.1 & .442 & .248 & 41.1 \\
& Base (0.6B)   & 58.0 & 26.3 & 32.4 & .479 & .249 & 41.5 \\
& LRAT (0.6B)   & 72.0 & 18.8 & 43.7 & .596 & .321 & 40.0 \\
\midrule
\multirow{3}{*}{\iter{} (0.6B)}
& SQ only
  & 72.0 & 19.1
  & 45.3\,\textcolor{red}{\scriptsize(+3.7\%)}
  & .624\,\textcolor{red}{\scriptsize(+4.7\%)}
  & .343\,\textcolor{red}{\scriptsize(+6.9\%)} & 39.9 \\
& MQ+SQ+PSQ
  & \textbf{79.3}\,\textcolor{red}{\scriptsize(+10.1\%)}\sigmark{*} & \textbf{17.6}
  & 45.7\,\textcolor{red}{\scriptsize(+4.6\%)}
  & .641\,\textcolor{red}{\scriptsize(+7.6\%)}
  & .342\,\textcolor{red}{\scriptsize(+6.5\%)} & 40.5 \\
& MQ+SQ+PSQ+PR (default)
  & 78.7\,\textcolor{red}{\scriptsize(+9.3\%)}\sigmark{*} & 18.2
  & \textbf{49.2}\,\textcolor{red}{\scriptsize(+12.6\%)}\sigmark{*}
  & \textbf{.661}\,\textcolor{red}{\scriptsize(+10.9\%)}
  & \textbf{.350}\,\textcolor{red}{\scriptsize(+9.0\%)} & \textbf{39.5} \\
\midrule
\multirow{3}{*}{\iter{} (4B)}
& SQ only & 73.3 & 18.3 & 46.0 & .644 & .348 & 39.5 \\
& MQ+SQ+PSQ & 78.7 & 18.5 & 48.3 & .637 & .349 & 39.5 \\
& MQ+SQ+PSQ+PR (default) & \textbf{80.3} & \textbf{16.7} & \textbf{51.2} & \textbf{.651} & \textbf{.370} & \textbf{39.4} \\
\bottomrule
\end{tabular}}
\end{table}

\section{Main Results}
We evaluate \iter{} in two settings. The matched-agent evaluation uses Tongyi-DeepResearch-30B for both trajectory collection and evaluation, while the cross-agent evaluation tests transfer to five unseen agent backbones.

\subsection{Matched-Agent Evaluation}
\label{sec:results:matched}

Table~\ref{tab:main} reports the matched-agent results. From these comparisons, we draw three findings about trajectory-relative training, interaction history, and retriever scale.

\finding{1}{Trajectory-relative training \revisiondel{already improves performance}\,\revisionadd{improves out-of-domain performance} with the current sub-query held fixed.}

To isolate the effect of trajectory-relative training, we compare LRAT with the current-sub-query-only \iter{} variant while holding the encoder backbone and query representation fixed. Both use Qwen3-Embedding-0.6B and receive only the current sub-query, but differ in how their training signals are constructed from agent trajectories. \revisiondel{On InfoSeek-Eval, \iter{} increases task success from 72.7 to 76.7 and reduces the average number of tool calls from 19.6 to 18.3.}\ \revisionadd{On InfoSeek-Eval, \iter{} matches LRAT at 72.0 task success, with comparable average tool calls (19.1 versus 18.8).} On BrowseComp-Plus, it increases evidence search recall from \revisiondel{.602 to .619}\,\revisionadd{.596 to .624} and task success from \revisiondel{43.4 to 43.7}\,\revisionadd{43.7 to 45.3}. Because the retrieval input is identical, these \revisionadd{out-of-domain }gains isolate the benefit of constructing supervision relative to the agent's position in the trajectory\revisionadd{, while the InfoSeek-Eval result shows that the benefit does not transfer uniformly across benchmarks}.

\finding{2}{Encoding the main question and previous sub-queries provides further gains.}

We next compare the query representations used to train and evaluate \iter{}. Adding the main question and previous sub-queries to the current-sub-query-only representation raises InfoSeek-Eval task success from \revisiondel{76.7 to 80.0}\,\revisionadd{72.0 to 79.3}. On BrowseComp-Plus, evidence search recall increases from \revisiondel{.619 to .636}\,\revisionadd{.624 to .641} and task success from \revisiondel{43.7 to 46.6}\,\revisionadd{45.3 to 45.7}. These components therefore provide \revisiondel{most of the BrowseComp-Plus improvement over LRAT}\,\revisionadd{the largest gain on InfoSeek-Eval and improve evidence retrieval on BrowseComp-Plus}. Further adding pre-search reasoning increases BrowseComp-Plus evidence search recall to \revisiondel{.648}\,\revisionadd{.661} and task success to \revisiondel{47.6}\,\revisionadd{49.2}, but lowers InfoSeek-Eval task success by \revisiondel{1.0}\,\revisionadd{0.6} point. Thus, the main question and previous sub-queries provide useful context beyond the current sub-query, while pre-search reasoning produces mixed gains across the two benchmarks.

\finding{3}{Increasing retriever scale \revisiondel{does not consistently improve performance}\,\revisionadd{generally improves task success}.}

\revisiondel{Scaling \iter{} from 0.6B to 4B produces mixed results. On BrowseComp-Plus, task success improves for the current-sub-query-only and default representations, but slightly decreases for the reasoning-augmented representation. On InfoSeek-Eval, the current-sub-query-only and default representations both drop by 3.0 points, while the reasoning-augmented representation remains unchanged. Evidence search recall is similarly mixed, improving only for the current-sub-query-only representation.}\ \revisionadd{Scaling \iter{} from 0.6B to 4B generally improves task success. All three query representations improve on BrowseComp-Plus; on InfoSeek-Eval, the default and SQ-only representations improve, whereas the MQ+SQ+PSQ representation decreases slightly. Evidence search recall remains mixed across representations.}

History conditioning remains beneficial at the larger scale. At 4B, the default representation outperforms the setting using only the current sub-query by \revisiondel{3.3}\,\revisionadd{7.0} points on InfoSeek-Eval and \revisiondel{2.7}\,\revisionadd{5.2} points on BrowseComp-Plus. Because LRAT does not provide a 4B model, these results characterize how \iter{} behaves when scaled from 0.6B to 4B rather than providing a same-scale comparison with LRAT. Increasing retriever scale therefore does not replace history conditioning, which remains beneficial even with the larger encoder.
   % Controlled Evaluation

\begin{table}[htbp]
\centering
\caption{Cross-agent evaluation of LRAT, AgentIR, and \iter{} across six agent backbones. SR = task success
rate; Search/Visit Recall = BrowseComp-Plus evidence search/visit recall; V/S =
visit-to-search recall ratio, a proxy for how effectively retrieved gold evidence
is converted into agent-visited evidence; Steps = mean tool calls. SR annotations use scale-matched
references: 0.6B \iter{} is relative to LRAT (0.6B), and 4B \iter{}
is relative to AgentIR (4B)
(\textcolor{red}{red}: increase; \textcolor{green!55!black}{green}: decrease);
bold marks the best value in each column within each retriever-scale block.
Stars indicate significant differences
under two-sided exact McNemar tests for the same agent backbone: 0.6B \iter{}
is compared with LRAT (0.6B), and 4B \iter{} with AgentIR
(4B): $^{*}p{<}.05$.}
\label{tab:backbones}
\small
\renewcommand{\arraystretch}{1.0}
\setlength{\tabcolsep}{3pt}
\resizebox{\textwidth}{!}{%
\begin{tabular}{l l c c c c c c c}
\toprule
\multirow{2}{*}{Agent Backbone} & \multirow{2}{*}{Retriever}
  & \multicolumn{2}{c}{\textbf{InfoSeek-Eval} (ID)}
  & \multicolumn{5}{c}{\textbf{BrowseComp-Plus} (OOD)} \\
\cmidrule(lr){3-4}\cmidrule(lr){5-9}
& & SR ($\uparrow$) & Steps ($\downarrow$)
  & SR ($\uparrow$) & Search Recall ($\uparrow$) & Visit Recall ($\uparrow$) & V/S (\%) ($\uparrow$) & Steps ($\downarrow$) \\
\midrule
\multirow{4}{*}{Tongyi-30B}
& LRAT (0.6B) & 72.0 & 18.8 & 43.7 & .596 & .321 & \textbf{53.8} & 40.0 \\
& \iter{} (0.6B)
  & \textbf{78.7}\,\textcolor{red}{\scriptsize(+9.3\%)}\sigmark{*} & \textbf{18.2}
  & \textbf{49.2}\,\textcolor{red}{\scriptsize(+12.6\%)}\sigmark{*}
  & \textbf{.661} & \textbf{.350} & 52.9 & \textbf{39.5} \\
\cmidrule(lr){2-9}
& AgentIR (4B) & 77.3 & 17.2 & 50.4 & \textbf{.724} & \textbf{.375} & 51.9 & \textbf{39.3} \\
& \iter{} (4B)
  & \textbf{80.3}\,\textcolor{red}{\scriptsize(+3.9\%)} & \textbf{16.7}
  & \textbf{51.2}\,\textcolor{red}{\scriptsize(+1.6\%)}
  & .651 & .370 & \textbf{56.9} & 39.4 \\
\midrule
\multirow{4}{*}{Qwen3.5-4B}
& LRAT (0.6B) & 70.7 & 14.2 & 25.5 & .452 & .265 & \textbf{58.7} & \textbf{24.5} \\
& \iter{} (0.6B)
  & \textbf{74.0}\,\textcolor{red}{\scriptsize(+4.7\%)} & \textbf{13.8}
  & \textbf{28.2}\,\textcolor{red}{\scriptsize(+10.6\%)}
  & \textbf{.472} & \textbf{.277} & \textbf{58.7} & 24.7 \\
\cmidrule(lr){2-9}
& AgentIR (4B) & 74.3 & \textbf{13.3} & \textbf{32.9} & \textbf{.568} & \textbf{.345} & 60.8 & 23.9 \\
& \iter{} (4B)
  & \textbf{75.3}\,\textcolor{red}{\scriptsize(+1.3\%)} & 13.8
  & 30.1\,\textcolor{green!55!black}{\scriptsize($-$8.5\%)}
  & .517 & .325 & \textbf{62.9} & \textbf{23.5} \\
\midrule
\multirow{4}{*}{Qwen3.5-9B}
& LRAT (0.6B) & 73.3 & 13.2 & 30.5 & .522 & .312 & 59.8 & \textbf{25.3} \\
& \iter{} (0.6B)
  & \textbf{77.3}\,\textcolor{red}{\scriptsize(+5.5\%)} & \textbf{12.9}
  & \textbf{34.9}\,\textcolor{red}{\scriptsize(+14.4\%)}\sigmark{*}
  & \textbf{.544} & \textbf{.344} & \textbf{63.3} & 25.5 \\
\cmidrule(lr){2-9}
& AgentIR (4B) & 76.7 & 12.1 & \textbf{39.0} & \textbf{.632} & \textbf{.411} & 65.1 & 25.1 \\
& \iter{} (4B)
  & \textbf{78.7}\,\textcolor{red}{\scriptsize(+2.6\%)} & \textbf{11.7}
  & 38.8\,\textcolor{green!55!black}{\scriptsize($-$0.5\%)}
  & .577 & .392 & \textbf{67.9} & \textbf{25.0} \\
\midrule
\multirow{4}{*}{Qwen3.5-27B}
& LRAT (0.6B) & 81.0 & 12.7 & 42.5 & .602 & .426 & 70.7 & \textbf{24.9} \\
& \iter{} (0.6B)
  & \textbf{84.0}\,\textcolor{red}{\scriptsize(+3.7\%)} & \textbf{11.8}
  & \textbf{44.9}\,\textcolor{red}{\scriptsize(+5.6\%)}
  & \textbf{.610} & \textbf{.446} & \textbf{73.2} & 25.1 \\
\cmidrule(lr){2-9}
& AgentIR (4B) & 81.3 & \textbf{11.5} & \textbf{52.4} & \textbf{.699} & \textbf{.518} & 74.1 & 25.0 \\
& \iter{} (4B)
  & \textbf{82.3}\,\textcolor{red}{\scriptsize(+1.2\%)} & 12.4
  & 51.3\,\textcolor{green!55!black}{\scriptsize($-$2.1\%)}
  & .658 & .497 & \textbf{75.6} & \textbf{24.7} \\
\midrule
\multirow{4}{*}{Qwen3.6-27B}
& LRAT (0.6B) & 71.0 & 13.9 & 29.0 & .378 & .189 & 50.0 & \textbf{17.2} \\
& \iter{} (0.6B)
  & \textbf{82.0}\,\textcolor{red}{\scriptsize(+15.5\%)}\sigmark{*} & \textbf{11.8}
  & \textbf{37.2}\,\textcolor{red}{\scriptsize(+28.3\%)}\sigmark{*}
  & \textbf{.510} & \textbf{.261} & \textbf{51.1} & 18.3 \\
\cmidrule(lr){2-9}
& AgentIR (4B) & 79.7 & \textbf{11.3} & 41.4 & \textbf{.613} & \textbf{.313} & 51.1 & 18.8 \\
& \iter{} (4B)
  & \textbf{80.3}\,\textcolor{red}{\scriptsize(+0.8\%)} & 12.3
  & \textbf{41.9}\,\textcolor{red}{\scriptsize(+1.2\%)}
  & .555 & .296 & \textbf{53.3} & \textbf{18.6} \\
\midrule
\multirow{4}{*}{gpt-oss-120B}
& LRAT (0.6B) & 68.0 & 10.4 & 35.2 & .543 & .296 & 54.5 & \textbf{19.1} \\
& \iter{} (0.6B)
  & \textbf{69.7}\,\textcolor{red}{\scriptsize(+2.5\%)} & \textbf{9.0}
  & \textbf{42.5}\,\textcolor{red}{\scriptsize(+20.7\%)}\sigmark{*}
  & \textbf{.611} & \textbf{.337} & \textbf{55.1} & 19.4 \\
\cmidrule(lr){2-9}
& AgentIR (4B) & \textbf{71.0} & 9.6 & \textbf{43.7} & \textbf{.694} & \textbf{.372} & 53.6 & 19.2 \\
& \iter{} (4B)
  & 69.3\,\textcolor{green!55!black}{\scriptsize($-$2.4\%)} & \textbf{9.4}
  & 43.6\,\textcolor{green!55!black}{\scriptsize($-$0.2\%)}
  & .629 & .359 & \textbf{57.1} & \textbf{18.5} \\
\bottomrule
\end{tabular}}
\end{table}

\subsection{Cross-Agent Evaluation}
\label{sec:results:generalization}

Table~\ref{tab:backbones} reports retriever performance across six agent backbones, including Tongyi-DeepResearch-30B, which generated the training trajectories. For each backbone, we hold the agent configuration fixed and change only the retriever. These comparisons yield three findings about cross-agent transfer, \revisiondel{pre-search reasoning}\ \revisionadd{retriever scaling across agent backbones}, and performance relative to AgentIR.

\finding{1}{\iter{} consistently outperforms LRAT across agent backbones.}

Across the six agent backbones, \revisiondel{the default \iter{} improves task success over LRAT in all 12 backbone--benchmark comparisons. The average gains are $+5.4$ points on InfoSeek-Eval and $+4.3$ points on BrowseComp-Plus, with seven gains statistically significant at $p{<}.05$.}\ \revisionadd{the 0.6B \iter{} consistently outperforms LRAT in all 12 backbone--benchmark comparisons, achieving average relative improvements of 6.9\% on InfoSeek-Eval and 15.4\% on BrowseComp-Plus.} On BrowseComp-Plus, \revisiondel{evidence search recall also improves for all six backbones, by 4.8 percentage points on average.}\ \revisionadd{evidence search recall and visit recall also improve for all six backbones, by 5.3 and 3.4 percentage points on average, respectively.} The benefit of \iter{} is therefore not limited to the agent that generated its training trajectories, but extends across three model families and agent scales ranging from 4B to 120B.

\finding{2}{\revisiondel{Pre-search reasoning transfers less reliably across agent backbones.}\ \revisionadd{Scaling \iter{} yields consistent out-of-domain gains across agent backbones.}}

\revisiondel{The matched-agent evaluation shows that adding pre-search reasoning can improve performance with Tongyi-DeepResearch-30B. Across the five unseen backbones, however, the reasoning-augmented representation performs worse than the default representation on InfoSeek-Eval in every case, on BrowseComp-Plus in four of five cases, and in evidence search recall in all five cases. Although it still outperforms LRAT in most unseen-agent comparisons, its gains are less consistent than those of the default representation.}

\revisionadd{Scaling \iter{} from 0.6B to 4B improves BrowseComp-Plus task success for all six agent backbones, by 3.3 points on average. Visit recall and the visit-to-search recall ratio also improve for every backbone, with the latter increasing from 59.1\% to 62.3\% on average, while search recall improves for five of six backbones. This consistency across agent backbones extends the matched-agent scaling result in Table~\ref{tab:main}, showing that the benefit is not specific to the agent used for trajectory collection. On InfoSeek-Eval, task success improves for three backbones and decreases for three, indicating that the cross-agent scaling benefit is concentrated on the out-of-domain setting.}

\revisiondel{One possible explanation is that pre-search reasoning is free-form and model-specific, so different agent backbones may express the same search intent differently. Because the reasoning-augmented representation is trained only with reasoning produced by Tongyi, it may transfer less reliably to unseen agents. These results reveal a transfer trade-off specific to pre-search reasoning: it can help with the matched agent but becomes less reliable when the agent backbone changes.}

\finding{3}{\revisiondel{AgentIR's higher evidence search recall does not reliably translate into evidence visits or task success.}\ \revisionadd{\iter{} remains competitive with AgentIR using implicit interaction signals.}}

\revisiondel{AgentIR performs strongly with Tongyi, but its search recall advantage transfers less reliably to unseen agents. In the size-matched 4B comparison, AgentIR achieves higher evidence search recall than the default \iter{} on four of five unseen backbones, whereas \iter{} achieves higher task success on all five. Against the smaller 0.6B \iter{}, AgentIR has higher search recall on three of five backbones, while \iter{} has higher visit recall on four of five and higher task success on all five. Thus, retrieving relevant evidence does not guarantee that the agent will consume it or use it successfully.}

\revisionadd{AgentIR is a strong reasoning-aware retriever. Its DR-Synth pipeline adds gold positive documents to the candidate pool and uses an LLM reranker to select positives and hard negatives~\cite{agentir2026}. This direct optimization of document relevance may explain AgentIR's high search recall. In contrast, \iter{} constructs training data directly from agent trajectories, using interaction signals already produced during trajectory collection that require no gold document labels or oracle-augmented reranking.}

\revisiondel{Unlike \iter{}, AgentIR relies on LLM-as-a-judge relevance labels rather than document-visit signals from agent trajectories~\cite{agentir2026}. These labels identify relevant documents but do not indicate which documents an agent will visit. Consequently, AgentIR may repeatedly rank relevant but unvisited documents highly, whereas \iter{} learns from the evidence agents actually consume. This difference may explain \iter{}'s stronger transfer in visit recall and task success.}

\revisionadd{This difference is reflected in how retrieved evidence translates into agent visits and task success. Although \iter{}'s search recall is 5.7 points lower than AgentIR's on average, the gap narrows to 1.6 points in visit recall. \iter{} achieves a higher visit-to-search recall ratio for every backbone, averaging 62.3\% versus 59.4\%. Despite its lower search recall, its end-to-end effectiveness is not correspondingly reduced: \iter{} achieves higher task success in 7 of the 12 backbone--benchmark comparisons.}

\revisionadd{Beyond the current offline setting, unlike AgentIR, \iter{} could in principle support continual online retriever updates by turning newly completed agent trajectories directly into training data.}
   % Cross-Agent Generalization

% Presentation IDs vs logged/RESULTS.md IDs -- keep this mapping when updating:
% paper i3 = logged i6 (docs), paper i4 = logged i5 (notes/DI),
% paper i5 = logged i7 (docs + notes),
% paper i6 = logged i8 (AgentIR format: PR + SQ),
% paper i7 = logged i9 (i2 + PR). Logs and RESULTS.md keep i8/i9 unchanged.

\begin{table}[t]
\centering
\caption{Query-input ablation with Tongyi. All \iter{} variants share the same
training setup; LRAT is included for reference. DI = document interpretation extracted from post-visit reasoning; \texttt{docs} = visited-document text;
other abbreviations as in Table~\ref{tab:main}. Recall = evidence search recall. Superscripts \(a\) and \(b\) mark significant differences from LRAT and
the default \revisiondel{i2}\ \revisionadd{i7} setting, respectively, using two-sided exact McNemar tests for SR
and paired \(t\)-tests for recall, with Bonferroni correction by metric and
anchor family at \(p<.05\).}
\label{tab:abl-input}
\small
\begin{tabular}{l l ccc}
\toprule
& \makecell[l]{Retriever input} & \makecell{InfoSeek-Eval\\SR} & \makecell{BCP\\SR}
& \makecell{BCP\\Recall} \\
\midrule
\textit{LRAT} & \textit{SQ}                    & \textit{72.0} & \textit{43.7} & \textit{.596} \\
i0 & SQ                                       & 72.0\sigmark{b} & 45.3 & .624\sigmark{b} \\
i1 & MQ, SQ                                   & 78.0 & 44.2\sigmark{b} & .636\sigmark{a} \\
i2 & MQ, SQ, PSQ                              & \textbf{79.3}\sigmark{a} & 45.7 & .641\sigmark{a} \\
\midrule
i3 & MQ, SQ, PSQ, docs                        & 75.0 & 42.0\sigmark{b} & .550\sigmark{ab} \\
i4 & MQ, SQ, PSQ, DI                          & 75.7 & 43.5\sigmark{b} & .594\sigmark{b} \\
i5 & MQ, SQ, PSQ, docs, DI                    & 75.0 & 42.7\sigmark{b} & .568\sigmark{b} \\
\midrule
i6 & SQ, PR                                   & 76.3 & 48.1 & \textbf{.668}\sigmark{a} \\
\textbf{i7} & MQ, SQ, PSQ, PR (default)       & 78.7\sigmark{a} & \textbf{49.2}\sigmark{a} & .661\sigmark{a} \\
\bottomrule
\end{tabular}
\end{table}

\begin{table}[t]
\centering
\caption{Negative-tier ablation of the default \iter{} configuration with
  Tongyi. Weights are ordered as
  $(w_{\text{red}},w_{\text{hard}},w_{\text{weak}})$. Notation follows Table 3.}
\label{tab:abl-neg}
\small
\setlength{\tabcolsep}{4pt}
\begin{tabular}{l ccc}
\toprule
Variant & \makecell{InfoSeek-Eval\\SR}
& \makecell{BCP\\SR} & \makecell{BCP\\Recall} \\
\midrule
\textit{LRAT}                  & \textit{72.0} & \textit{43.7} & \textit{.596} \\
default $(3.0,1.0,0.3)$ & 78.7\sigmark{a} & \textbf{49.2}\sigmark{a} & \textbf{.661}\sigmark{a} \\
\midrule
w/o redundancy            & 79.0\sigmark{a} & 43.5\sigmark{b} & .600\sigmark{b} \\
w/o hard                  & 77.0 & 46.0 & .658\sigmark{a} \\
\midrule
uniform $(1,1,1)$          & 77.0 & 45.3 & .633\sigmark{ab} \\
stronger red. $(5,1,0.2)$  & \textbf{80.3}\sigmark{a} & 46.9 & .653\sigmark{a} \\
\bottomrule
\end{tabular}
\end{table}

\section{Ablation Analysis}

We analyze the two main design choices in the default \iter{} configuration: the history-conditioned query representation and trajectory-relative negative supervision. We vary each component separately while holding the other fixed.

\subsection{Query Representation}

We first vary the query representation while holding the training examples, negative tiers, and training recipe fixed. Each representation is used during both training and evaluation. Table~\ref{tab:abl-input} compares eight variants. Variants i0--i2 progressively add the main question and previous sub-queries; i3--i5 incorporate visited documents, post-visit document interpretations, or both; and i6--i7 examine pre-search reasoning. Variant i6 follows the representation used by AgentIR~\cite{agentir2026}, while i7 adds pre-search reasoning to \revisiondel{the default representation}\ \revisionadd{i2 and serves as the default representation}.

\finding{1}{The main question and previous sub-queries provide complementary \revisiondel{gains}\ \revisionadd{context}.}

\revisiondel{Adding the main question (i0 $\rightarrow$ i1) leaves InfoSeek-Eval task success unchanged at 76.7, but improves BrowseComp-Plus task success from 43.7 to 45.4 and evidence search recall from .619 to .629. Further adding previous sub-queries (i1 $\rightarrow$ i2) raises InfoSeek-Eval task success to 80.0, BrowseComp-Plus task success to 46.6, and evidence search recall to .636.}\ \revisionadd{Adding the main question (i0 $\rightarrow$ i1) raises InfoSeek-Eval task success from 72.0 to 78.0 and evidence search recall from .624 to .636, but lowers BrowseComp-Plus task success from 45.3 to 44.2. Further adding previous sub-queries (i1 $\rightarrow$ i2) improves all three metrics, raising them to 79.3, 45.7, and .641, respectively.} The main question preserves the overall task, while previous sub-queries indicate which search directions have already been explored. Their combination therefore provides the strongest representation among i0--i2.

\finding{2}{Explicitly encoding visited content hurts retrieval.}

Adding visited documents (i3), post-visit interpretations (i4), or both (i5) lowers \revisiondel{BrowseComp-Plus task success to 41.1--42.7 and evidence search recall to .541--.579}\ \revisionadd{InfoSeek-Eval task success to 75.0--75.7, BrowseComp-Plus task success to 42.0--43.5, and evidence search recall to .550--.594}, with all three variants performing worse than i2.

One possible explanation is that visited-document text conflicts with the training signal: it places consumed evidence in the query representation while the same documents may serve as redundancy negatives that the retriever is trained to rank lower. Document interpretations summarize what the agent learned more abstractly, which may explain why i4 performs better than i3 on BrowseComp-Plus. Nevertheless, all three variants remain below i2. This result indicates that representing previously explored directions through sub-queries is more effective than directly encoding consumed evidence or its interpretation.

\finding{3}{\revisiondel{Pre-search reasoning provides agent- and benchmark-specific gains.}\ \revisionadd{Pre-search reasoning primarily improves out-of-domain retrieval.}}

\revisiondel{Adding pre-search reasoning improves BrowseComp-Plus performance with Tongyi-DeepResearch-30B but does not improve InfoSeek-Eval. As shown in Section~\ref{sec:results:generalization}, these gains also transfer less consistently to unseen agent backbones. Pre-search reasoning can therefore provide useful context for the agent that produced the training trajectories, but it is less reliable than the structured default representation across benchmarks and agents.}\ \revisionadd{Adding pre-search reasoning consistently improves BrowseComp-Plus task success and evidence recall, whether paired only with the current sub-query or added to the structured history. The full i7 representation provides the strongest out-of-domain performance while remaining competitive on InfoSeek-Eval, motivating its use as the default. These results suggest that pre-search reasoning primarily benefits out-of-domain retrieval.}

\subsection{Negative Supervision}

We next examine the construction and weighting of trajectory-relative negatives. Table~\ref{tab:abl-neg} compares the effects of removing negative tiers and varying their loss weights.

\finding{1}{Redundancy negatives provide the strongest \revisionadd{out-of-domain }trajectory-relative signal.}

Removing redundancy negatives has the largest \revisionadd{out-of-domain }effect, reducing BrowseComp-Plus task success from \revisiondel{46.6 to 40.8}\ \revisionadd{49.2 to 43.5} and evidence search recall from \revisiondel{.636 to .561}\ \revisionadd{.661 to .600}, while InfoSeek-Eval task success \revisiondel{decreases by only 0.7 points}\ \revisionadd{increases by 0.3 point}. This result captures the central distinction of interaction-aware retrieval: a document may remain relevant to the current sub-query while providing little additional value after its information has been consumed. The large \revisionadd{out-of-domain }drop shows that learning to recognize this changing utility is the main contribution of trajectory-relative negative supervision.

\finding{2}{Hard negatives provide complementary relevance supervision.}

Removing hard negatives reduces task success from \revisiondel{80.0 to 79.3}\ \revisionadd{78.7 to 77.0} on InfoSeek-Eval and from \revisiondel{46.6 to 45.8}\ \revisionadd{49.2 to 46.0} on BrowseComp-Plus. Although these effects are smaller than those of redundancy negatives, the declines on both benchmarks show that conventional relevance supervision remains useful. Hard negatives help distinguish irrelevant documents from useful evidence, complementing the history-dependent signal provided by redundancy negatives.

\finding{3}{\revisiondel{Negative tiers require calibrated weighting.}\ \revisionadd{Negative-tier weighting trades off in-domain and out-of-domain performance.}}

Uniform weighting $(1,1,1)$ assigns uncertain weak negatives the same weight as redundancy negatives. This reduces BrowseComp-Plus task success from \revisiondel{46.6 to 43.9}\ \revisionadd{49.2 to 45.3} and evidence search recall from \revisiondel{.636 to .608}\ \revisionadd{.661 to .633}. The default weighting $(3,1,0.3)$ instead reflects the strength of the interaction evidence: redundancy negatives receive the largest weight, visited but irrelevant documents retain the standard weight, and unvisited documents are discounted. It provides the best \revisiondel{overall}\ \revisionadd{out-of-domain} balance, reaching task success of \revisiondel{80.0}\ \revisionadd{78.7} on InfoSeek-Eval and \revisiondel{46.6}\ \revisionadd{49.2} on BrowseComp-Plus.

Further increasing the separation with weights $(5,1,0.2)$ does not improve BrowseComp-Plus \revisionadd{(46.9)} and \revisiondel{reduces InfoSeek-Eval task success to 75.0}\ \revisionadd{improves InfoSeek-Eval task success to 80.3}. The negative tiers should therefore not be treated uniformly\revisiondel{, but the redundancy signal should not dominate the training objective}\revisionadd{.} The default weighting balances strong evidence of redundancy against the uncertainty of unvisited documents\revisionadd{, leading to more robust transfer beyond the training distribution}.
   % Analysis and Ablations
\section{Conclusion}
\label{sec:conclusion}

Deep-research agents search cumulatively: each sub-query and document visit changes what evidence remains useful. We introduced \iter{}, an interaction-aware dense retriever that models a document's marginal gain using a history-conditioned query representation and trajectory-relative supervision.

Across two benchmarks and six agent backbones, \iter{} consistently outperforms LRAT and \revisiondel{shows stronger cross-agent transfer than AgentIR, achieving higher task success for all five unseen backbones on both benchmarks}\ \revisionadd{remains competitive with AgentIR at the matched 4B scale, while it does not require sophisticated and expensive training data creation}. Our analyses show that \revisiondel{the main question and previous sub-queries provide robust retrieval context, whereas document visits are more effective as supervision than as query content, and free-form pre-search reasoning transfers less reliably}\ \revisionadd{the main question, previous sub-queries, and pre-search reasoning provide complementary retrieval context, whereas document visits are more effective as supervision than as query content. Scaling \iter{} further improves performance across agent backbones.}

More broadly, these results suggest that agentic retrieval should optimize progress through a search trajectory, rather than relevance at an isolated step. Agent trajectories provide a useful starting point, but fully capturing the agent's evolving information state remains an open problem. Important questions include how to represent what an agent has learned, transfer interaction signals across agents, and jointly optimize retrieval and search decisions. Addressing them could shift retrieval from finding relevant evidence to finding the evidence most useful next for the agent.
   % Conclusion
\section*{Acknowledgments}
We thank \href{https://kid-22.github.io/}{\textcolor{blue}{Sunhao Dai}} and
\href{https://yuqi-zhou.github.io/}{\textcolor{blue}{Yuqi Zhou}} for helpful discussions.
% \clearpage
\bibliography{iter}

\end{document}